\input{aipcheck}

\documentclass[
    ,final            
  ]
  {aipproc}

\layoutstyle{8x11single}

\def\sun{\hbox{$\odot$}}

\begin{document}

\title{GeV photon absorption in cosmologically evolving quasar environments}


\author{A. Reimer}{
  address={Stanford University, HEPL \& KIPAC, Stanford, CA 94305, USA}
}

\begin{abstract}
The question of a possible redshift-dependence of $\gamma$-ray absorption due to $\gamma\gamma$-pair production of jet photons in the accretion disk and BLR radiation field in strong-line 
quasars is investigated. For this relevant aspects of cosmological black hole and 
quasar evolution are applied to the expected pair production opacity of GeV-photons in those sources.
I demonstrate that for positively evolving and non-evolving accretion rates over cosmological time, detectable
$\gamma$-ray optical depths originating within the AGN system will show a redshift-dependence in the LAT energy range with larger opacity
from sources at higher redshifts. 
This introduces ambiguities in the interpretation of spectral absorption features, and complicates approaches for estimating the evolution of the 
extragalactic background light by probing the $\gamma$-ray horizon when using blazars whose external photon fields are involved in 
$\gamma$-ray production.
\end{abstract}

\maketitle


\section{Motivation}

\vspace*{-.1cm}

The extragalactic background light (EBL) is the accumulated light in the optical/UV to infrared energy 
band resulting from structure formation and evolution. 
Since direct measurements suffer from large systematic uncertainties due to contamination of the data by the strong foreground, indirect methods through the absorption of $\gamma$-rays in the process of photon-photon pair production of extragalactic $\gamma$-ray sources (e.g. active galactic nuclei (AGN), GRBs) in the EBL seem promising. Current EBL models \cite[e.g.,][]{Primackxx,Kneiske04} indicate that spectral absorption features from $\gamma$-ray sources at redshift $z\geq 0.5$ are expected
in the LAT energy range (0.02-300GeV).
This offers the opportunity for GLAST to constrain the evolution of the EBL from high-energy measurements. 
The proposed methods trace the evolution of the horizon of $\gamma$-rays that have been emitted by extragalactic sources \cite[e.g.,][]{Kneiske04,Chen04} by e.g., determining the e-folding cutoff energy \cite[''Fazio-Stecker relation'';][]{Fazio70}. 
It implicitly assumes that any redshift-dependence of observed absorption features in a large source sample is the unique signature of the absorption occurring during propagation to Earth. Additional $\gamma$-ray absorption is, however, possible already within the source system.
The blazar class contains
a sufficient high number of $\gamma$-ray loud, bright, high-redshift sources,
and are therefore considered a promising source class for a statistically meaningful analysis to constrain EBL evolution.
Here I will concentrate on $\gamma$-ray loud quasars that show strong emission lines. Those are known as high-luminosity AGN in the $\gamma$-ray band, and thus have a marked probability of providing sufficient photon statistics to search for absorption breaks in a spectral analysis. In this source
class signatures of accretion disk radiation and emission lines from the cloud region have been regularly observed. Those radiation fields 
serve as target for $\gamma$-ray production, but at the same time constrain the spectrum in the >10 GeV regime by photon absorption through $\gamma\gamma$-pair production.
In the following I consider $\gamma$-ray absorption in the radiation field of an accretion disk (assumed here to follow the cool, optically thick blackbody solution of \cite{Shakura73}), and of the broad-line region (BLR)
\cite[see e.g.][]{Donea03}. 
A description of
the radiation field characteristics and calculations of this "local" $\gamma$-ray opacity can be found in \cite{Reimer07}.
The key step of this work is the application of cosmological black hole (BH) and quasar evolution to the expected "local" pair production opacity $\tau_{\gamma\gamma}$ of $\gamma$-ray photons in AGN.
  


\section{Application of black hole accretion rate evolution and results}

\vspace*{-.1cm}

With the availability of large AGN data archives enormous advances in BH demographic were made. The discovered 
remarkable similarity of the cosmic history of star formation and BH accretion rate density supports the vastly accepted picture of a joint BH--host galaxy evolution \cite[e.g.,][]{Marconi04,Lapi06,Netzer07}. The X-ray quasar luminosity functions evolution revealed the number density of fainter AGN to peak at lower redshifts than that of brighter ones, with implications
for the evolution of the BH mass function. The derived picture of ``anti-hierarchical'' BH growth via accretion has meanwhile been 
confirmed by many further studies.  
The resulting evolution of BH accretion rates, also recently confirmed by a direct analysis based on $\sim10^4$ type-I AGN of the SDSS \cite{Netzer07}, impacts the $\gamma$-ray quasar population where "intrinsic local"
$\gamma$-ray absorption in accretion disk and BLR radiation fields is potentially important. Three models for the evolution of the cosmic BH accretion rate (Fig.~1a) are applied to the calculations of the $\gamma$-ray opacity in the local AGN radiation fields from the accretion disk and BLR (see figure caption). Fig.~1a shows the
redshift-dependent BH growth and accretion rates with higher rates at larger redshifts.
The calculations are detailed in \cite{Reimer07}.

\begin{figure}
  \includegraphics[height=.2\textheight]{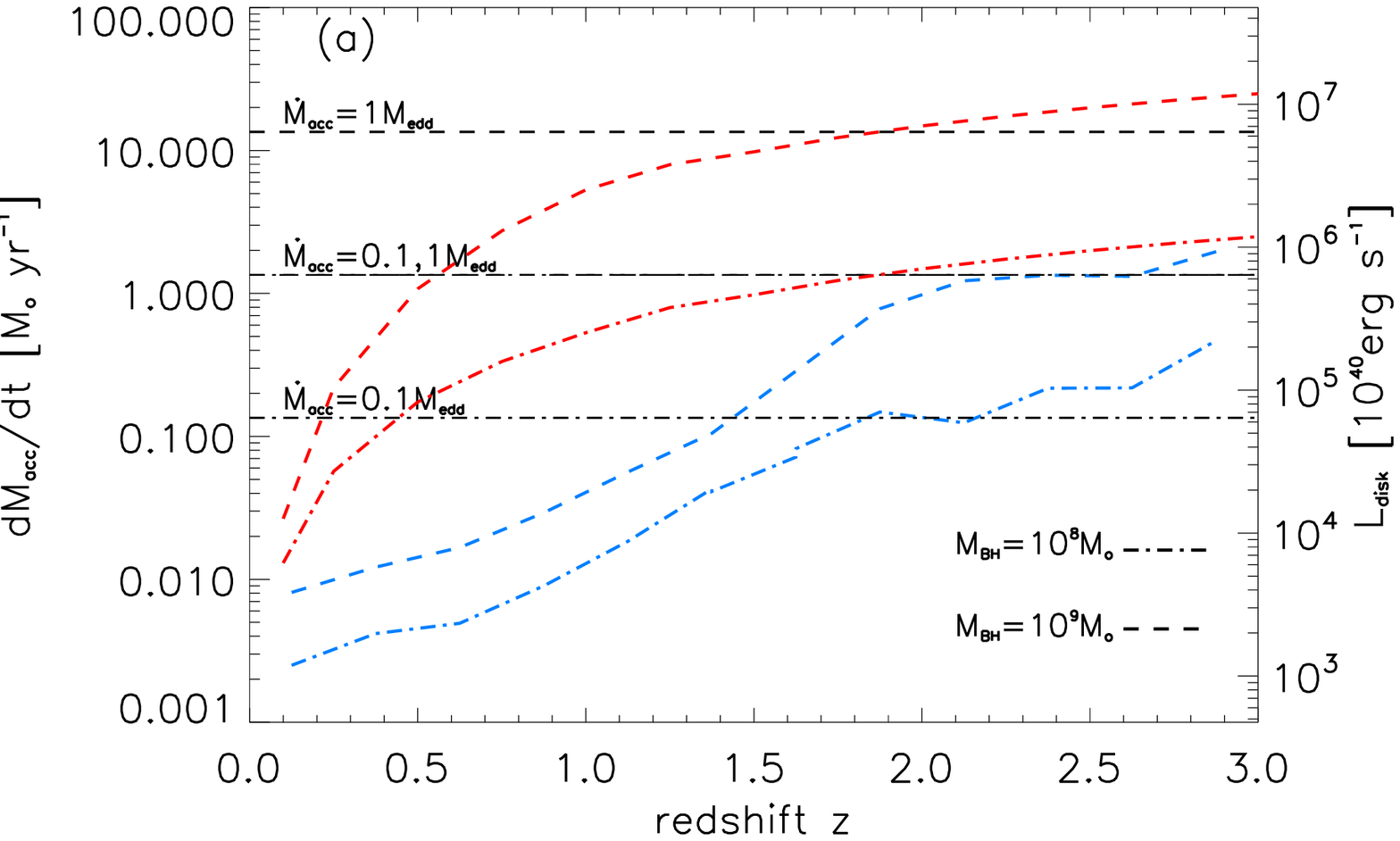}
\hfill
  \includegraphics[height=.2\textheight]{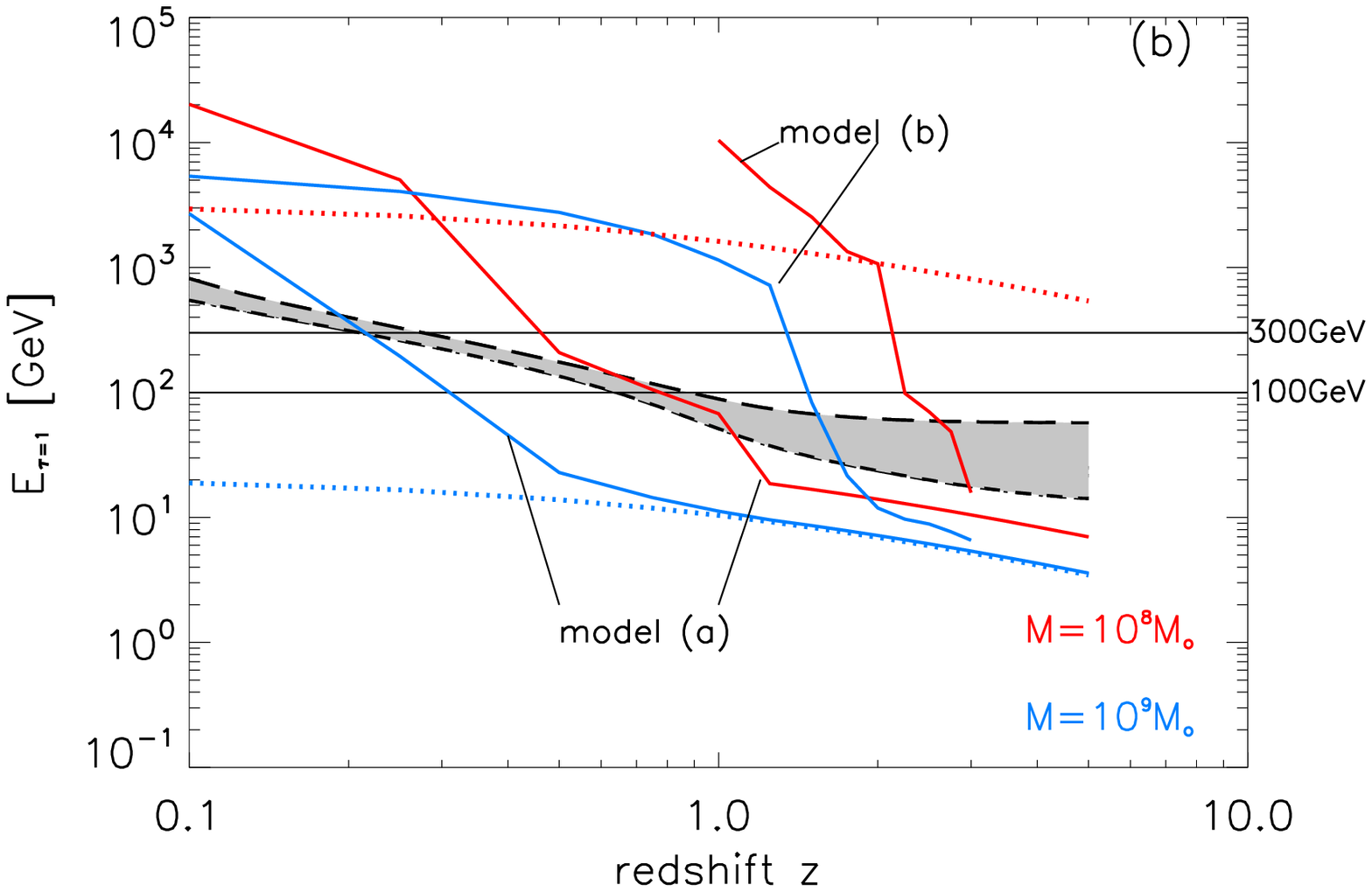}
\caption{
{\bf Left:} Evolution of BH mass accretion rates. Red (upper) curves: Results from the Netzer \& Trakhtenbrot \cite{Netzer07} analysis for $z\leq0.75$, complemented with a mild evolution at larger redshifts \citep{Lapi06}. Blue (lower) curves: Marconi et al. \cite{Marconi04} model. Black (straight) lines: no evolution with $\dot M_{\rm acc}=0.1-1M_{\rm edd}$. Dashed curves are for M$_{\rm BH}=10^8$M$_{\sun}$, dashed-dotted lines for M$_{\rm BH}=10^9$M$_{\sun}$. {\bf Right:}
E-folding cutoff energy $E(\tau_{\gamma\gamma}=1,z)$ versus redshift ($\gamma$-ray horizon) of jet $\gamma$-rays interacting with 
accretion disk and BLR photons with accretion rates following the Netzer \& Trakhtenbrot \cite{Netzer07} analysis, complemented by the Lapi et al \cite{Lapi06} model where only modest evolution is proposed at redshifts $>1$ (lower solid 
 colored curves; evolution model (a)) 
and the ``anti-hierarchical'' BH growth model of Marconi et al. \cite{Marconi04} (upper solid colored curves; evolution model (b)), and for non-evolving accretion rates (dotted curves
 for $\dot M_{\rm acc}=0.1$ (red; upper curve) and 1 (blue; lower curve)), compared to the Fazio-Stecker presentation of various evolutionary EBL models
 of \cite{Kneiske04} (grey shaded area). 
All models show a redshift-dependence of $E(\tau_{\gamma\gamma}=1)$.
\vspace*{-.4cm}
}
\end{figure}



Measurements of the e-folding cutoff energy for a large sample of extragalactic $\gamma$-ray sources may probe the evolution of the EBL unambiguously, if either $\gamma$-ray absorption local to the sources is negligible, or not redshift-dependent. For strong-line AGN I found:
\begin{itemize}

\vspace*{-.1cm}

\item If the $\gamma$-ray emission region is sufficiently close to the BLR, mandatory for $\gamma$-ray production on external photon fields, local $\gamma$-ray absorption features in quasar spectra have to be expected at $E(1+z)>$ several tens of GeV.
$\gamma$-ray absorption in LAT's energy range occurs mostly in the increasing part of the $\tau_{\gamma\gamma}$(E) function near threshold.
 \item BH demography reveals BH growth and corresponding accretion rates to be redshift-dependent with higher rates at larger redshifts. Correspondingly, the critical energy $E(\tau_{\gamma\gamma}=1)$ due to local absorption in quasar disk and BLR radiation fields decreases with redshift, similar to Fazio-Stecker's relation for EBL absorption (see Fig.~1b).
\item  Even for no evolution of disk accretion rates, $E(\tau_{\gamma\gamma}=1)$ decreases with redshift for AGN of a given BH mass (see Fig.~1b).
It results from the interplay of local absorption near pair production threshold and cosmological energy red-shifting.
\item  Any observed redshift-dependence of absorption features in strong-line quasars, which are prone to local $\gamma$-ray absorption, can therefore {\bf not} serve as a secure signature for absorption occurring in the EBL radiation field.
\item ``Naked jet sources'' (i.e. AGN without strong optical/UV radiation fields close to the $\gamma$-ray emission site) form a rather unbiased sample for studying the EBL evolution 
by probing the $\gamma$-ray horizon with GLAST's LAT.

\vspace*{-.2cm}

\end{itemize}





\bibliographystyle{aipproc}   


\IfFileExists{\jobname.bbl}{}
 {\typeout{}
  \typeout{******************************************}
  \typeout{** Please run "bibtex \jobname" to optain}
  \typeout{** the bibliography and then re-run LaTeX}
  \typeout{** twice to fix the references!}
  \typeout{******************************************}
  \typeout{}
 }



\end{document}